\documentclass{article}
\usepackage[english]{babel}
\usepackage[letterpaper,top=2cm,bottom=2cm,left=3cm,right=3cm,marginparwidth=1.75cm]{geometry}
\usepackage{amsmath,amsfonts,amsthm,authblk}
\usepackage{graphicx}
\usepackage{xcolor}
\usepackage[colorlinks=true, allcolors=blue]{hyperref}
\newcommand{\p}{\partial}

\usepackage{hyperref}

\title{A Continuum Model for Morphology Formation from Interacting Ternary Mixtures: Simulation Study of the Formation and Growth of Patterns}
\author[$\ddagger$]{Rainey Lyons \footnote{rainey.lyons@kau.se}}
\author[$\star$]{Stela Andrea Muntean}
\author[$\dagger$]{Emilio N. M. Cirillo} 
\author[$\ddagger$]{Adrian Muntean}

\affil[$\dagger$]{Department of Basic and Applied Sciences for Engineering (SBAI), Sapienza University of Rome, Italy}

\affil[$\ddagger$]{Department of Mathematics and Computer Science, Karlstad University, Sweden
 }

\affil[$\star$]{Department of Engineering and Physics, Karlstad University, Sweden
 }

 \usepackage{color}
\begin{document}

\maketitle

\begin{abstract}
 Our interest lies in exploring the ability  of a coupled nonlocal system of two quasilinear parabolic partial differential equations to produce phase separation patterns. The obtained patterns are referred  here as morphologies.  Our target system is derived in the literature as the rigorous hydrodynamic limit of a suitably scaled  interacting particle system of Blume--Capel--type 
driven by Kawasaki dynamics.  The system describes in a rather implicit way the interaction within a ternary mixture that is the macroscopic counterpart of a mix of two populations of interacting solutes in the presence of a background solvent. Our discussion is based on the qualitative behavior of numerical simulations of  finite volume approximations of smooth solutions to our system and their quantitative postprocessing in terms of two indicators (correlation and structure factor calculations). Our results show many similar features compared to what one knows at the level of the stochastic Blume--Capel dynamics with three interacting species. The properties of the obtained morphologies (shape, connectivity, and so on) can play a key role in, e.g., the design of the active layer for efficient organic solar cells.\\[0.25cm]
\newline
{\textbf{PACS Subject Classification:}} 64.75.Qr,  64.75.St, 02.60.Lj
\newline
\end{abstract}

\section{Introduction}

We are interested in exploring the capability of interacting mixtures of populations of particles, typically two different solutes mixed within a solvent, to allow for phase separation when evaporating the solvent.  Such situations are rather typical in modern materials science. In particular, we have in mind two specific materials science applications: the formation of internal morphologies for organic solar cells (cf. e.g. \cite{hoppe2004organic}); and the formation of rubber--based zones in acrylate environments related to the design of thin adhesive bands (cf. e.g. \cite{creton2016rubber}). 
The evaporation process plays a key role in both of these contexts. The presence of such a non--equilibrium process makes this scenario different compared to the more classical phase separation settings arising in crystal growth, metallurgy, and so on, normally treated by Allen--Cahn--type  or Cahn--Hilliard--type equations. 
 Essentially, by controlling the evaporation mechanism one has the possibility to freeze a local equilibrium, i.e. one has the possibility to actively select specific morphologies as the end configuration. We refer the reader to the recent works \cite{Andrea_EPJ,Andrea_PhysRevE, Mario}, where this setting has been explored using Monte Carlo simulations for suitable lattice--based models.  With such final and stable morphologies at hand (see e.g. Figure~1 in \cite{Andrea_EPJ}), one can then ask key questions about their efficiency with respect to the effective charge transport in the case of the organic solar cells or their efficiency as a global (effective) measure of stickiness in the case of adhesive bands. 

To this end, it is crucial to have sufficiently rich models posed at the continuum level that inherit, from the interacting particle systems, the ability to produce physically meaningful phase separation.  This would set the stage for process and shape optimization operations to reach optimal macroscopic transport and reaction properties, mediated by optimal morphology shapes. However, quite interestingly, in spite of huge efforts in the statistical mechanics and applied mathematics communities, deriving rigorously the corresponding hydrodynamic limit equations for mixtures of interacting particle systems having the possibility of evaporation at one of the boundaries is currently out of reach. The main technical difficulties seems to arise because of the evaporation component of the process. Therefore, we lay foundation in this manuscript by looking at a simplified case where
the evaporation is not taken into account. 
The problem remains interesting and relevant, with a rich phenomenology, 
for the presence of three coexisting phases undergoing 
the separation process.

In \cite{Andrea_EPJ,Andrea_PhysRevE}, and follow--up papers,
a generalized version
of the 
nearest neighbor
Blume--Capel model in $2D$ with 
Kawasaki dynamics powered with the Metropolis algorithm was employed as working tool 
to simulate the phase behaviour of a
ternary mixture upon evaporation of one of its components.
Here 
we are genuinely interested in devising the corresponding 
hydrodynamics limit in absence of evaporation.  
The components
with spin $+1$, $-1$, and $0$ in 
the Blume--Capel dynamics correspond
to two distinct solute phases and, respectively, to solvent\footnote{When handling the active layer for an organic solar cell, the mixture is composed of electron--acceptor, electron--donor and solvent molecules, respectively.}. 
In particular, in \cite{Andrea_PhysRevE}
it was used 
as governing Hamiltonian 
the usual nearest neighbor Blume--Capel Hamiltonian
with zero chemical potential and magnetic field, see \cite{CO1996}.

Although we are using working techniques 
from \cite{Presutti} to investigate the situation, we are unaware of 
the corresponding hydrodynamic limit equations arising from this precise 
structure of the stochastic model. 
However, 
for a  modified Hamiltonian structure involving Kac potentials 
much is known. Specifically, the authors of \cite{Marra} rigorously 
identify the continuum limit model for both the Kawasaki 
and Glauber dynamics in the infinite volume case
(so, the evaporation processes is not taken into account). We refer the 
reader also to \cite{GLP_1999,Giacomin_Lebowitz_Marra} for related 
discussions on the same class of models.

The phase ordering dynamics of the two--field model derived 
in \cite{Marra} will be the object of our study.
We remark that studying equilibrium and non--equilibrium properties in 
connection with phase transitions for statistical 
mechanics models with multiple degenerate ground states
is a natural extension of the theories developed for
two--state systems, such as the paradigmatic Ising model, which,
independently from possible real-world applications, has a great 
theoretical value. 
In particular, when dynamical phenomena are concerned, new 
interesting behaviors emerge. 
We refer, for instance, 
to \cite{CO1996,cirillo2013relaxation,cirillo2017sum} 
for the study of metastability and 
to \cite{PAB1997} for the phase ordering 
problem. 

Phase ordering dynamics of Ising--like two--phase mixtures has been 
widely studied and it is well understood; compare e.g.  
\cite{bray2002theory,binder1991,gunton1983}. 
The interval of times in which the domain morphology does not change 
is characterized by the dynamical scaling of the correlation 
function $G(\vec{x},t)$, where $\vec{x}$ and $t$ are the space and 
time variable, respectively. This can be expressed, in the isotropic case, 
by saying that $G(\vec{x},t)=g(|\vec{x}|/L(t))$, where $L(t)$ is the 
time dependent length scale, which is interpreted 
as the typical size of the growing domains. 
Both for conservative and non--conservative dynamics, 
the length scale is characterized by a power law $L(t)\sim t^\alpha$. Simulations suggest the approximate value of exponent $\alpha$ is $1/2$ for the non--conservative and $1/3$ for 
the conservative dynamics. While there is evidence that the scaling hypothesis holds in many situations, 
it has only been shown to be true for certain models; see, for instance,   \cite{amar1990diffusion,bray1990universal,coniglio1989multiscaling,derrida1991scale} as well as \cite{Charlie_Paquay_McLeish} for a closely related context.

For
systems with multiple phases the understanding of the phase separation 
processes is far from being complete. For some specific models, such as the clock 
and the Potts model with conservative dynamics, 
it is well established that the late times growth exponent
is $1/2$, see, e.g., \cite{PAB1997} and references therein. Note though that  some controversies have been more recently pointed out in 
\cite{IdBFCLP}.
What concerns those dynamics conserving the order parameter, not many studies
are available. We mention, again, \cite{PAB1997} for the case of the clock 
model. Here the authors have shown a crossover between exponents 
$1/4$ and $1/3$. 
A particular version of the nearest neighbor Blume--Capel model was investigated in \cite{Andrea_PhysRevE}. The reported simulations indicate that the $1/3$ growth behavior can be reached, but results are not yet conclusive.

The main focus of this work is to explore,
in the context of the 
two--field model derived in \cite{Marra} and
by means of numerical simulations, three main ideas: 
\begin{itemize}
\item[--]
to which extent the two-field continuum model reported in  \cite{Marra} for the Kawasaki dynamics is able to produce phase separation; 
\item[--]
how the obtained patterns (called here morphologies) relate to what is known for the classical Ising model in 2D and to what we have observed ourselves in our quantitative study \cite{Andrea_PhysRevE} for the three--component 
mixture driven by 
the standard Blume--Capel model 
\begin{equation}\label{Eq_HamAndrea}
H(\sigma) = \frac{1}{2} \sum_{\substack{x\neq x'\in V:\\|x-x'| =1}} 
             [\sigma(x)-\sigma(x')]^2,
\end{equation}
where $V\subset\mathbb{Z}^2$ is a finite square, 
with or without the evaporation of one of the phases; 
\item[--]
to which extent the scaling hypothesis holds in this setting.  
\end{itemize}

This work is organized as follows: in Section~\ref{model} we define 
the model and 
describe the finite volume scheme we use to compute decent approximations 
of solutions to the target model. 
In the main section of the manuscript, Section \ref{simulation}, 
we report the observed morphologies for selected parameter regimes borrowed from \cite{Andrea_PhysRevE} so that direct comparisons between the two settings can be done. The analysis of the growth of morphologies on intermediate and long time scales is done in  Section \ref{Sec_DomainGrowth} using common measures applied to interacting particle systems modified so they are applicable to problems posed at the continuum level. We illustrate numerically that such methods are able to approximate the characteristic length of the domains. Although this is rather common knowledge in the statistical mechanics--type exploration of patterns obtained e.g. with lattice models, our type of quantitative analysis of patterns seems to be novel in the context of phase--field models of Cahn--Hilliard type. In the same section, we present evidence explaining to which extent model \eqref{Eq_MainModel} satisfies the scaling hypothesis. Finally, we close the paper with a conclusion and outlook Section \ref{outlook}.

\section{The model}\label{model}

Let $\emptyset\neq\Omega\subset{\mathbb{R}}^2$ be a bounded open set with 
smooth boundary (e.g., $\partial\Omega$ Lipschitz) describing the spatial 
domain for the processes to take place. Let $T>0$ represent the final time 
of the overall dynamics. The time $T$ takes a large positive value, which in 
the current setting\footnote{If the evaporation process is included in the 
model, then one can relate the value of the final time of the overall 
process $T$ to the amount of evaporated solvent.} 
is picked arbitrarily. 
We set $\beta>0$, $m_0:(0,T)\times \bar\Omega\to [-1,+1]$, 
and $\phi_0:(0,T)\times \bar\Omega\to [0,+1]$. 
We are interested in the following 
two-field continuum model:
find the pair $(m,\phi)$ solving the system 
\begin{equation}\label{Eq_MainModel}
\left\lbrace
\begin{split}
    \p_t m &= \nabla \cdot \left[\nabla m - 2 \beta (\phi -m^2 ) (\nabla J * m) \right] \mbox{ in } (0,T)\times\Omega\\
    \p_t \phi &= \nabla \cdot \left[ \nabla \phi - 2 \beta m (1 - \phi) (\nabla J * m) \right]\mbox{ in } (0,T)\times\Omega
\end{split}\right., 
\end{equation}
where 
$J\in C_+^2(\mathbb{R}^2)$, symmetric, compactly supported, 
and $\int_{\mathbb{R}^2}J(r)dr=1$, 
together with the initial data
\begin{equation}
m(t=0)=m_0  \mbox{ and } \phi(t=0)=\phi_0 \mbox{ in } \bar\Omega
\end{equation}
and with periodic boundary conditions. 
We refer to this system as problem $P(\Omega)$.  

The problem $P({\mathbb{R}}^2)$ 
is derived 
in \cite{Marra} 
as
the rigorous hydrodynamic limit 
of the Kawasaki dynamics 
with inverse temperature $\beta$
for the Blume--Capel model with Kac potential 
with range of 
interaction $\gamma^{-1}$, magnetic field $h_1$, and chemical potential
$h_2$,
whose Hamiltonian 
in a finite square $V\subset\mathbb{Z}^2$ 
is 
\begin{equation}\label{Eq_HamMarra}
H_\gamma (\sigma) 
= 
\frac{1}{2} \sum_{\substack{x \neq x' \in V}} 
J_\gamma (x-x')[\sigma(x)-\sigma(x')]^2 
-\sum_{x\in V}h_1 \sigma(x) 
-\sum_{x\in V}h_2 \sigma^2(x), 
\end{equation}
where $J_\gamma: \mathbb{R}^2\to \mathbb{R}$ 
is such that 
\begin{equation}
\label{geigamma}
J_\gamma(r)=\gamma^2 J(\gamma r)
\end{equation}
for all $r\in \mathbb{R}^2$.

We are studying here the framework offered by $P(\Omega)$ and have 
in view the situation when one can think of the diameter of the 
set $\Omega$, say $\text{diam}(\Omega)$, to be large. However, we are 
not currently concerned with the case $\text{diam}(\Omega)\to\infty$.

In \cite{Marra}, $m$ is referred to as magnetization, while $\phi$ is called concentration. Their precise physical meaning seems to be best understood at the level of the stochastic dynamics. In other words, if the spin variable at the site $i\in\mathbb{Z}^2$ is denoted by $\sigma(i)$, then for some set  $\Lambda\subset \mathbb{Z}^2$ the empirical measures $\sum_{i\in\Lambda}\sigma(i)$ and $\sum_{i\in\Lambda}\sigma^2(i)$ would recover in a certain way\footnote{To obtain the structure of our continuum model, a scaling similar to \cite{Giacomin} is employed in the arguments presented in \cite{Marra}.} 
(see Section 3 in \cite{Marra}) as weak--star limits the 
densities $m$ and $\phi$, respectively. In our interpretation of the model for a subset $A \subset \Omega$, $\int_{A}m(t,x) dx$ represents the net spin in the set $A$ and  $\int_{A} (1-\phi(t,x)) dx$ represents the solvent fraction in $A$. In our visualizations for $\phi$ shown in Section \ref{simulation}, we will make use of the color red to point out the  time--space distribution of the solvent fraction. If the values taken by the image of $m$ are close to $-1$, and respectively $+1$, then we point out the  the  time--space distribution of the other two competing phases. We visualize these phases with yellow and blue colors, respectively.

\subsection{Finite Volume Scheme}
\label{scheme}
We begin by discretizing the domain with a mesh of sizes $\Delta x , \Delta y$. We denote the uniformly spaced nodes of the mesh by the pair $(x_i,y_j)$ and we denote the mesh cells by $\Lambda_{i,j} := [x_i -\frac12 \Delta x, x_i +\frac12 \Delta x) \times [y_j -\frac12 \Delta y, y_j +\frac12 \Delta y)  $ with the natural modifications for the boundary. The initial pair $(m_0, \phi_0)$ is made discrete with the approximations
\[ m^0_{i,j} := \frac{1}{\vert \Lambda_{i,j}\vert}\int_{\Lambda_{i,j}} m_0(x,y) dxdy \quad \text{ and } \quad \phi^0_{i,j} :=\frac{1}{\vert \Lambda_{i,j}\vert}\int_{\Lambda_{i,j}} \phi_0(x,y) dxdy .\]
We then approximate the solution to \eqref{Eq_MainModel} by a fully explicit finite--volume (difference) scheme given by 

\begin{equation}\label{Eq_Scheme1}
\left \lbrace
\begin{split}
    \frac{m^{k+1}_{i,j}-m^k_{i,j}}{\Delta t} &= \frac{1}{\Delta x^2}D^2_i[m^k_{i,j}] +\frac{1}{\Delta y^2}D^2_j[m^k_{i,j}]  \nonumber\\
    &- \frac{\beta}{\Delta x} D^1_i[(\phi^k_{i,j} -(m^k_{i,j})^2) \Tilde{J}^k_{x,i,j}] - \frac{\beta}{\Delta y} D^1_j[(\phi^k_{i,j} -(m^k_{i,j})^2) \Tilde{J}^k_{y,i,j} ] \\
    \frac{\phi^{k+1}_{i,j}- \phi^k_{i,j}}{\Delta t} &= \frac{1}{\Delta x^2}D^2_i[\phi^k_{i,j}] +\frac{1}{\Delta y^2}D^2_j[\phi^k_{i,j}]  \nonumber\\
    &- \frac{\beta}{\Delta x} D^1_i[m^k_{i,j}(1-\phi^k_{i,j}) \Tilde{J}^k_{x,i,j}] - \frac{\beta}{\Delta y} D^1_j[m^k_{i,j}(1-\phi^k_{i,j}) \Tilde{J}^k_{y,i,j} ]
\end{split} \right. ,
\end{equation}
where
\[D^2_l[f_l] := f_{l+1}-2f_{l} +f_{l-1}, \quad D^1_l[f_l] := f_{l+1} - f_{l-1},   \]
and $\Tilde{J}^k_{x,i,j}$ (respectively, $\Tilde{J}^k_{y,i,j}$) denotes the approximation of $\p_x J * m^k (x_i,y_j)$ (respectively, $\p_y J * m^k (x_i,y_j)$) using a fast Fourier transform method similar to \cite{TiwariKumaretal_2021_FastAccurateApproximation} with modifications for the periodic domain. We note that the periodic boundary conditions are implemented implicitly during the calculations. Since the scheme here is fully explicit, we make use of a Courant–Friedrichs–Lewy (CFL) condition similar to standard conditions for advection--diffusion processes, $\frac{\beta \Delta t}{ \Delta x \Delta y} < \frac{1}{4}$. We do not claim this choice of CFL condition is sharp, but it has held up through numerical experiments. Further rigorous analysis of the scheme studying its convergence and stability will illuminate more optimal conditions of relevance particularly when the parameter $\beta$ varies (i.e. when there are temperature changes in the system). 
There are many choices of numerical schemes for such equations. Since we are not interested in the particular geometry of the domain, we opt for the case where $\Omega$ is a square with periodic boundary conditions. Therefore, the choice of finite volume schemes is natural due to the conservative nature of the model. We also have in mind to eventually follow existing literature where schemes of similar flavor are used in the study of Cahn--Hillard equations (see e.g. \cite{kim2004conservative,kim2004conservativeTernary,cummings2018modeling}).

\section{Observed Morphologies}
\label{simulation}
We use the finite difference scheme described above to simulate the phase separation process according to the dynamics of \eqref{Eq_MainModel}. In all simulations, the initial conditions are chosen in a random manner such that $m^0_{i,j} \in \{-1,0,1\}$ with $\phi^0_{i,j} = |m^0_{i,j}|$. The proportion of positive spin ($m =1$) and negative spin ($m=-1$) particles is kept equal in each simulation, while the proportion of zero spin solvent particles (sometimes denoted by $c_0$) varies over the ranges 0\%--80\%. In all simulations, we take the parameter $\beta = 10$ and $J$ is taken to be a standard `bump' function such that $\int_{\mathbb{R}^2}J(r)dr=1$.

\begin{figure}
    \centering
    \includegraphics[scale = 0.6]{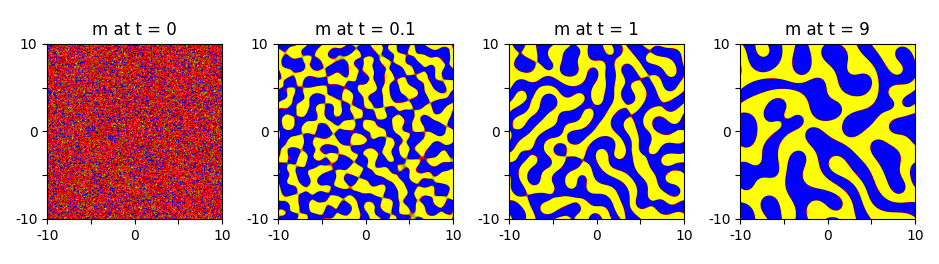}
    \includegraphics[scale = 0.6]{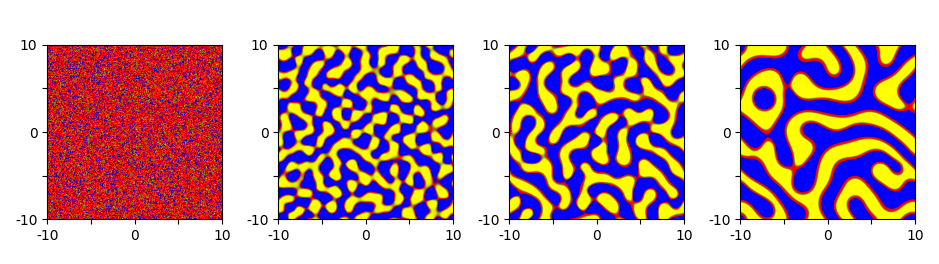}
    \includegraphics[scale = 0.6]{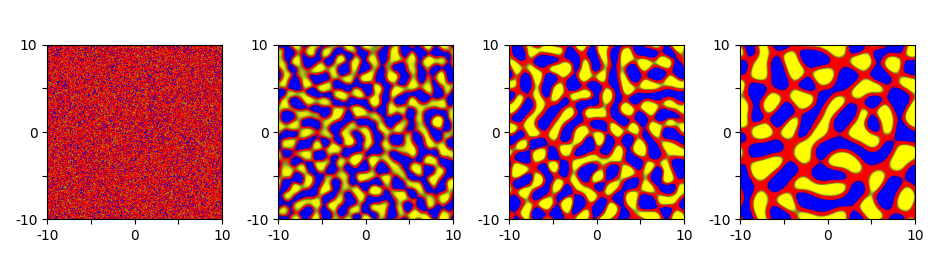}
    \includegraphics[scale = 0.6]{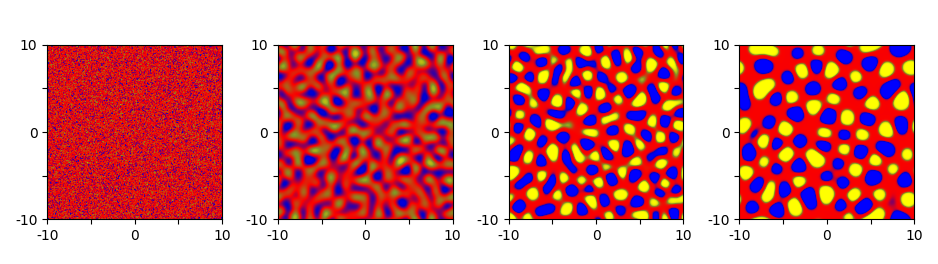}
    \includegraphics[scale = 0.6]{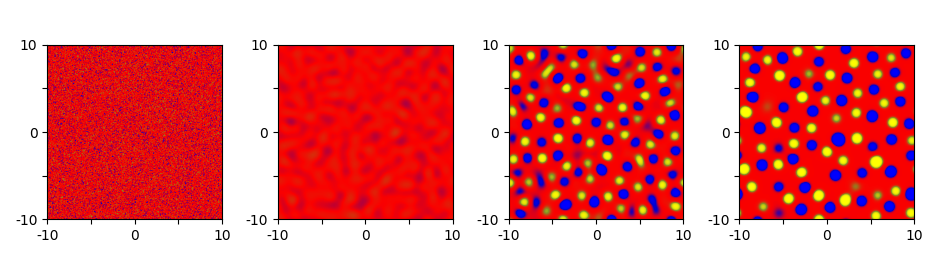}
    \caption{Morphologies produced by $m$ on a domain $\Omega := [-10,10]^2$ with 512 nodes in both the $x$ and $y$ axes. From top to bottom, each row represents an initial solvent concentration ($c_0$) of 0, 0.2, 0.4, 0.6, and 0.8, respectively. Regions where $m$ is positive are colored blue, negative regions are colored yellow, and regions $m$ is near zero are colored red. }
    \label{fig_m_Morphologies}
\end{figure}

\begin{figure}
    \centering
    \includegraphics[scale = 0.6]{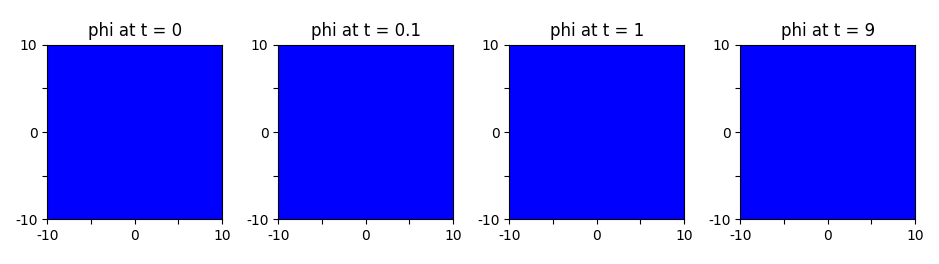}
    \includegraphics[scale = 0.6]{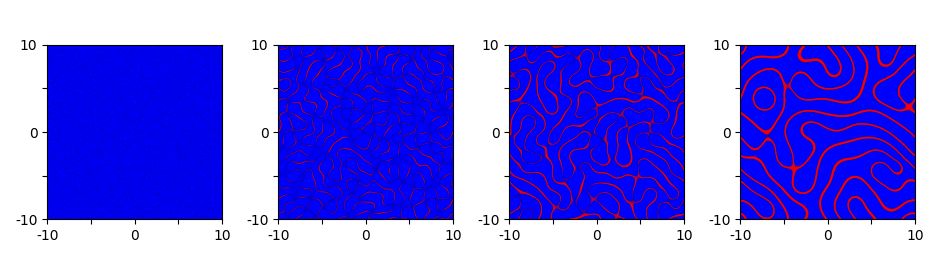}
    \includegraphics[scale = 0.6]{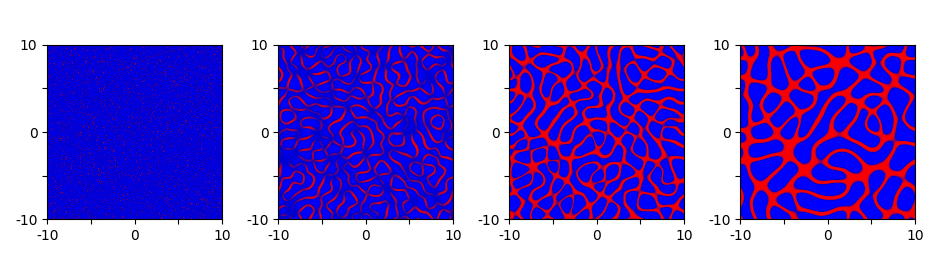}
    \includegraphics[scale = 0.6]{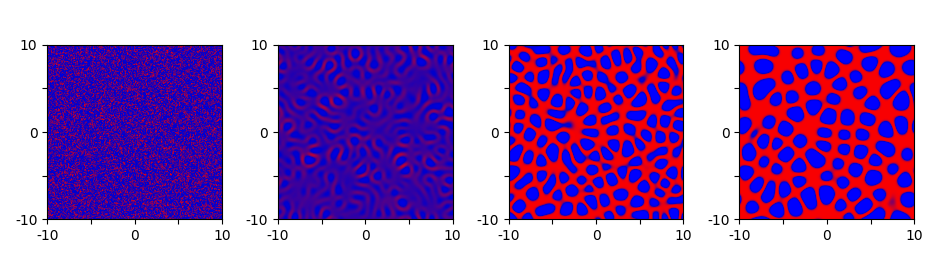}
    \includegraphics[scale = 0.6]{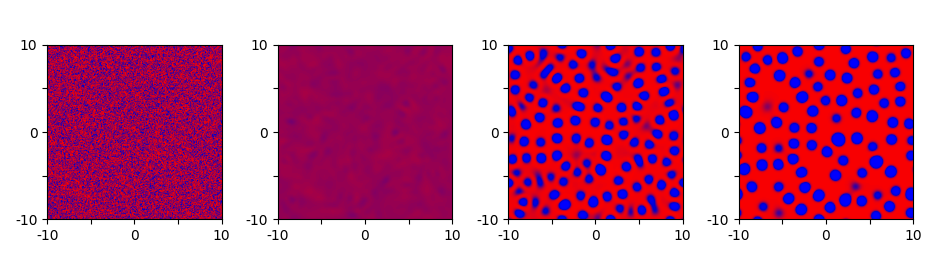}
    \caption{Morphologies produced by $\phi$ on a domain $\Omega := [-10,10]^2$ with 512 nodes in both the $x$ and $y$ axes. From top to bottom, each row represents an initial solvent concentration ($c_0$) of 0, 0.2, 0.4, 0.6, and 0.8, respectively. Regions where $\phi$ is near one are colored blue whereas regions near zero are colored red.}
    \label{fig_phi_Morphologies}
\end{figure}

\begin{figure}
    \centering
    \includegraphics[scale = 0.57]{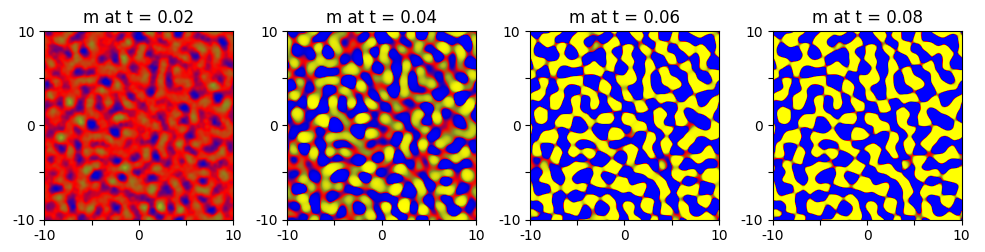}
    \includegraphics[scale = 0.57]{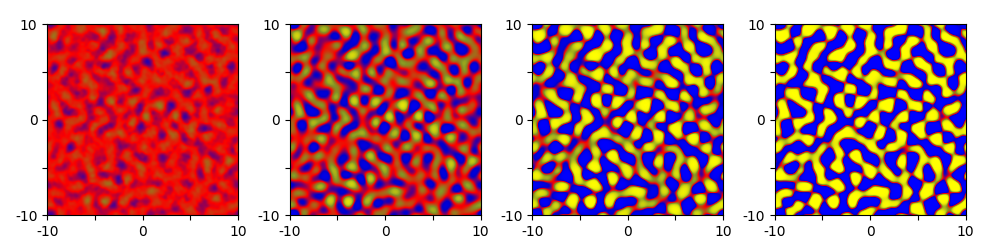}
    \includegraphics[scale = 0.57]{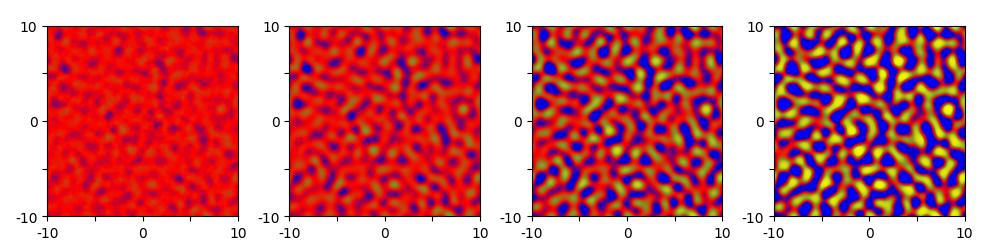}
    \includegraphics[scale = 0.57]{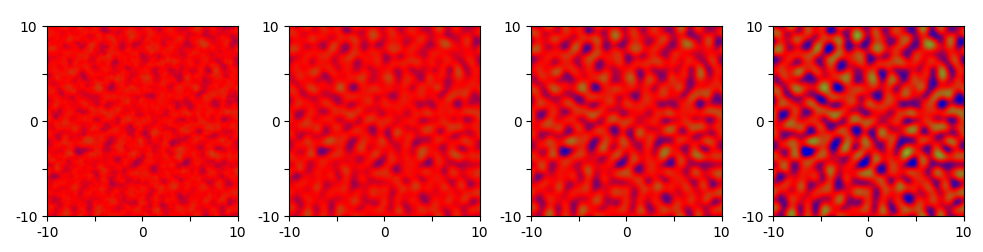}
    \includegraphics[scale = 0.57]{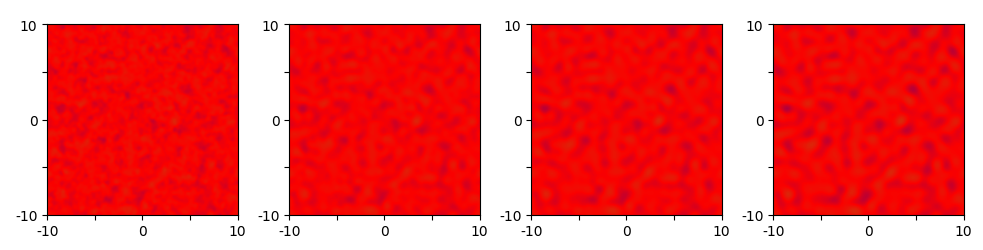}
    \caption{Morphologies produced by $m$ on a domain $\Omega := [-10,10]^2$ with 512 nodes in both the $x$ and $y$ axes. From top to bottom, each row represents an initial solvent concentration ($c_0$) of 0, 0.2, 0.4, 0.6, and 0.8, respectively. 
The color scale is as in Figure~\ref{fig_m_Morphologies}.}

    \label{fig_m_MorphologiesEarlyTime}
\end{figure}

\begin{figure}
    \centering
    \includegraphics[scale = 0.57]{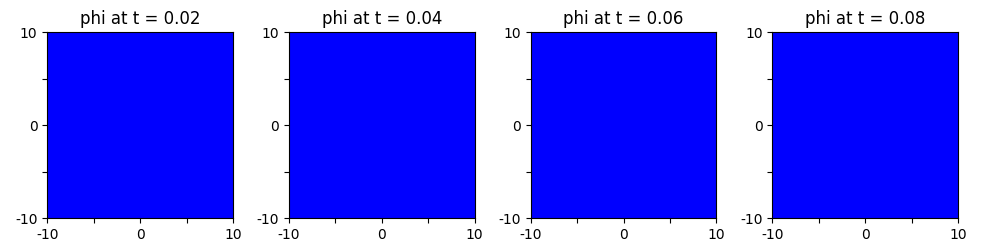}
    \includegraphics[scale = 0.57]{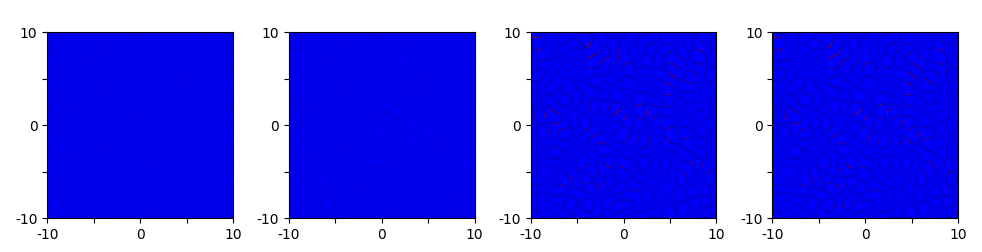}
    \includegraphics[scale = 0.57]{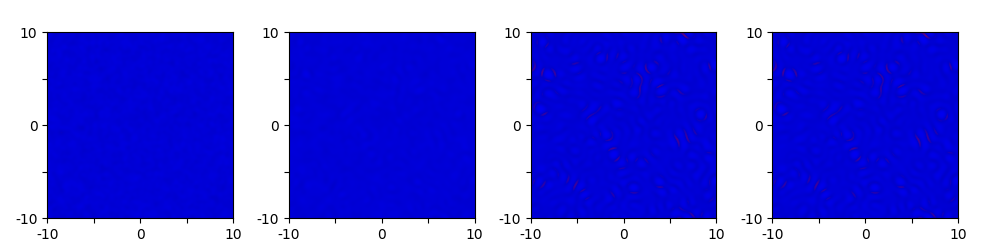}
    \includegraphics[scale = 0.57]{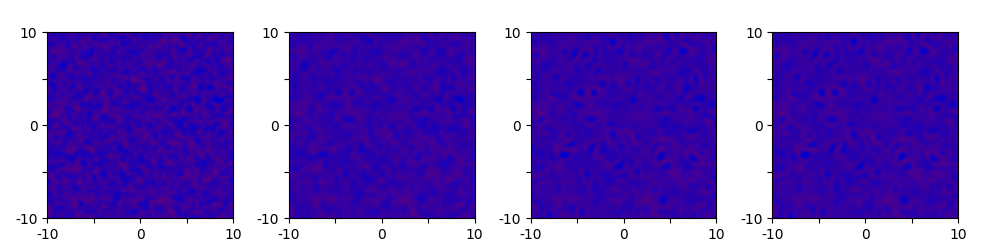}
    \includegraphics[scale = 0.57]{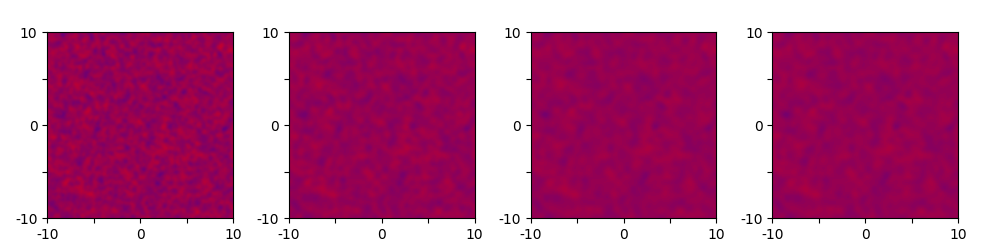}
    \caption{Morphologies produced by $\phi$ on a domain $\Omega := [-10,10]^2$ with 512 nodes in both the $x$ and $y$ axes. From top to bottom, each row represents an initial solvent concentration ($c_0$) of 0, 0.2, 0.4, 0.6, and 0.8, respectively. 
The color scale is as in Figure~\ref{fig_phi_Morphologies}.}

    \label{fig_phi_MorphologiesEarlyTime}
\end{figure}

In this section, we present the heat maps of the solutions $m$ and $\phi$ at different times and different solvent concentration levels in 
Figures~\ref{fig_m_Morphologies} and \ref{fig_phi_Morphologies}, respectively. Interestingly, one can see that if the solvent concentration $c_0$ is close to zero, then the produced bi--connected morphologies resemble to what one would obtain when simulating numerically the Ising model. The resemblance to the Ising model is best seen in Section \ref{large_scale}, where we study the domain growth at large characteristic timescales. The growth exponents that we obtain in the case $c_0=0$ 
are around $1/3$ as for the Ising model in $2D$, see, e.g.,
\cite{bray2002theory,binder1991,gunton1983}.

If the level of solvent increases, then the connectivity 
in the system decreases. Finally, a different type of morphology arises
(ball--like structures). This happens when the solvent concentration 
is so large (approximatively larger than $0.5$) that the solvent 
component percolates in the lattice: in this regime a unique connected 
solvent component is observed in the lattice \cite{grimmet}. In particular, looking at our simulation results shown in 
Figures~\ref{fig:Boxes}, \ref{fig:SmallTimeStructureFactor}, 
and \ref{fig_StructureFactorT9}, we see that indeed the growth rate of morphologies is decreasing with increasing $c_0$.  A solvent fraction at the level $c_0=0.4$ seems to offer some kind of separating threshold:  If $c_0$ lies below $0.4$, then both the morphologies and their growth rates are somewhat similar to what we would expect from the classical Ising scenario; If $c_0$ takes values greater than $0.4$, then the morphologies and their growth rates resemble more to what we would expect from standard percolation scenarios.

In the next section, we will see  that the indicators of the characteristic length scale oscillate during early times. To observe how these patterns initially form, we also present the heat maps of the simulation at early times in 
Figures~\ref{fig_m_MorphologiesEarlyTime} and \ref{fig_phi_MorphologiesEarlyTime}. For these early time simulations, one can see that neither the magnetization function, $m$, or the concentration function, $\phi$, are well organized. This demonstrates that there are at least two different time regimes under consideration in the continuum model. One at the early times before the morphologies are well formed and one at later times after the phases become separated. Multiple time regimes have been observed both experimentally and theoretically \cite{bray2002theory}. One also sees the beginnings of the morphologies for small concentrations (e.g. $c_0 = 0.2$) while for larger concentrations (e.g. $c_0 = 0.8$), the mixture is still cloud like in resemblance. This behavior seems to indicate different concentrations require  different amounts of time for phases to separate into robust morphologies. As one can see from 
Figure~\ref{fig_phi_MorphologiesEarlyTime}, the solvent concentration increasingly slows down the formation of morphologies.

\subsection{Analysis on Domain Growth}\label{Sec_DomainGrowth}

In this section, we analyse the domain growth of the simulations presented in Section \ref{simulation}. Common measures of the characteristic length of the domains include the two--point correlation function and the structure factor \cite{bray2002theory}. Since we are interested in a continuous setting, we present the continuous formulation of these metrics here. Their discrete counterparts can be found in \cite{Andrea_PhysRevE}; see also \cite{Gonnella,Gan}. The two--point correlation function is given by the equation
\[G(t,s) = \frac{1}{|\Omega|}\int_\Omega m(t,x,y) \, m(t,x+s_x,y+s_y) dxdy,\]
where $s=(s_x,s_y)$ and $|\Omega|$ denotes the volume of the set $\Omega$. In particular, the horizontal and vertical two--point correlation functions are defined as $G_x(t,s_x) = G(t,(s_x,0))$ and $G_y(t,s_y) =G(t,(0,s_y))$, respectively. These functions generally have a maximum at $s=0$, decrease below some cut--off value where the function begins to oscillate before returning back to the maximum (due to the periodic boundary conditions on the model). We can then estimate the horizontal and vertical diameters by calculating the horizontal intercept of $G_\alpha(t,\cdot)$ ($\alpha \in \{x,y\})$ using standard methods.

The structure factor of $m(t,\cdot)$ is defined as 
\[C(t,(k_x,k_y)) = \frac{1}{|\Omega |} \left| \int_\Omega m(t,x,y) e^{i(k_x x + k_y y)} dxdy \right|^2,\]
where $(k_x,k_y) \in [-\pi, \pi]^2.$ Notice the quantity inside the absolute value is the two--dimensional Fourier transform of $m(t,\cdot)$. We then estimate the horizontal and vertical diameters of the domains by calculating
\[R_\alpha (t) = \frac{\int_{[-\pi,\pi]^2} 
C(t,(k_x,k_y))
dk_x dk_y}{\int_{[-\pi,\pi]^2} |k_\alpha|
C(t,(k_x,k_y))
dk_x dk_y},\]
where $\alpha \in \{x,y\}$.

\begin{figure}[h]
    \centering
    \includegraphics[scale = 0.45]{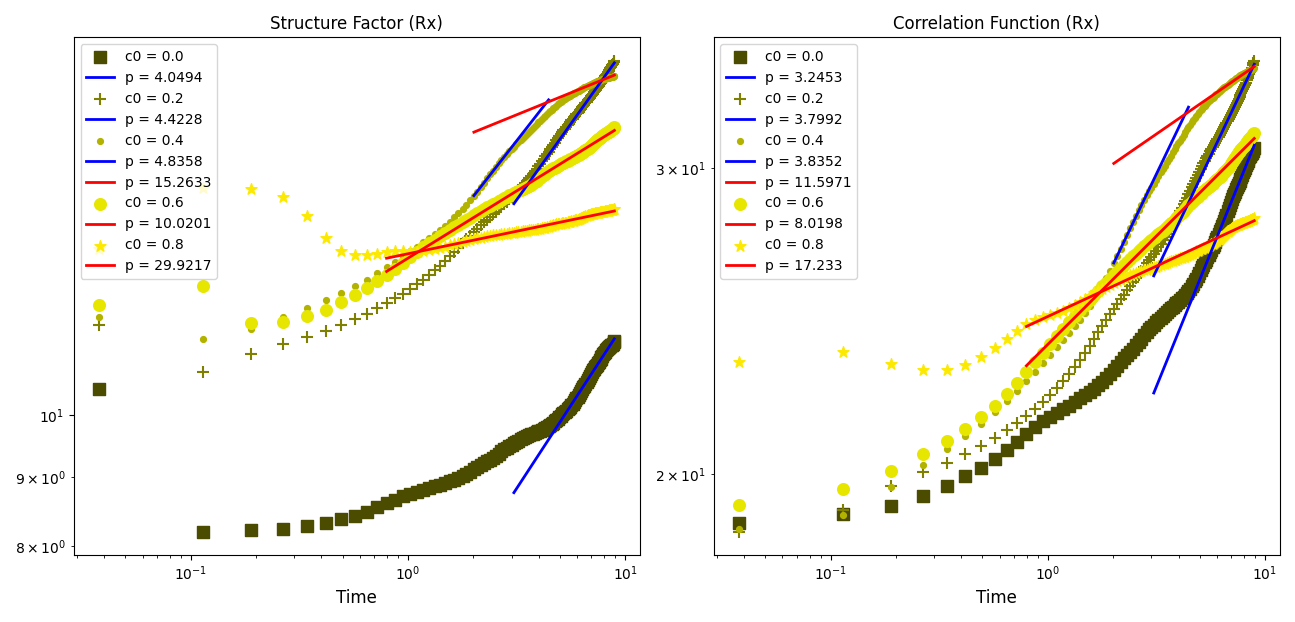}
    \caption{Structure factor (left) and correlation function (right) calculations for the simulations presented in Figure~\ref{fig_m_Morphologies} during the time interval $(0,9]$. Each marker represents $4,000$ time steps in the finite volume scheme. Lines of the form $y = Ct^{1/p}$ are plotted to demonstrate the approximate growth law of each concentration. The value of $p$ of each line is places in the legend underneath its corresponding solvent ratio level.}
    \label{fig_StructureFactorT9}
\end{figure}

\begin{figure}
    \centering
    \includegraphics[scale = 0.5]{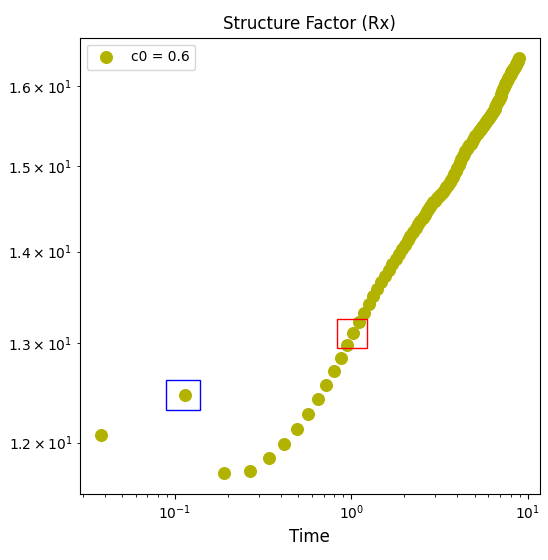}
    \includegraphics[scale = 0.5]{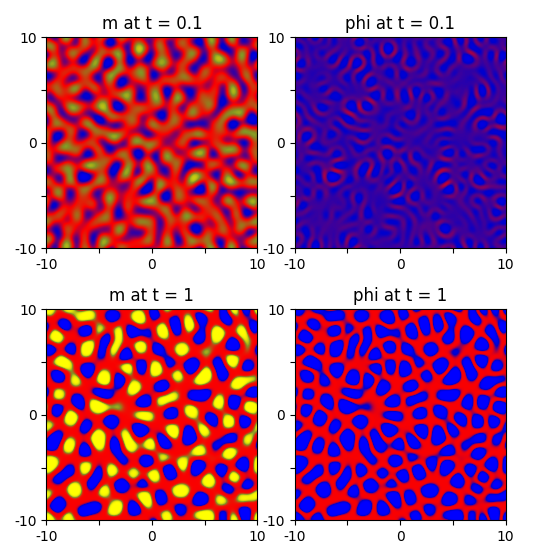}
    \caption{Demonstration of the different time regimes. Notice at early times (top/blue box), the $m$ and $\phi$ fields have not yet manifested well formed morphologies and so the growth of the structure factor does not fit a power law structure. On the other hand, for later times (bottom/red box), the growth is more consistent with a power law.}
    \label{fig:Boxes}
\end{figure}

In Figure~\ref{fig_StructureFactorT9}, we use the above indicators $G(\cdot,\cdot)$ and $C(\cdot,\cdot)$ to plot the domain sizes in time for all concentrations presented in Figures~\ref{fig_m_Morphologies}--\ref{fig_phi_MorphologiesEarlyTime}. In these plots, we observe an early time oscillatory behaviour and a later time linear (in a log scaling) growth which corresponds to the power laws of the scaling hypothesis. We observe in Figure~\ref{fig:Boxes}, that these two time regimes correspond to the time intervals where the domains are ill formed and well formed, respectfully. We analyse these two situations in more detail in the next section.

\subsection{Domain Growth at Intermediate Time Scales}\label{intermediate_scale}
\begin{figure}
    \centering
    \includegraphics[scale = 0.4]{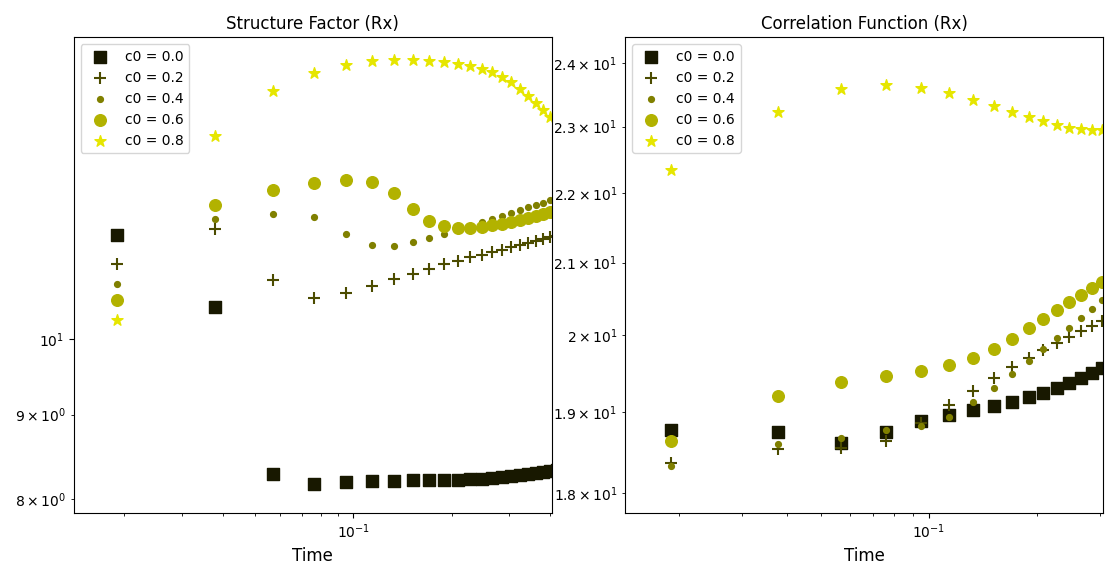}
    \caption{A zoomed in view of Figure~\ref{fig_StructureFactorT9} at early times. Here, each marker represents 1,000 time steps in the finite volume scheme. Note how each concentration level has different peaks indicating the dynamics are slowed by the solvent.}
    \label{fig:SmallTimeStructureFactor}
\end{figure}

In this section, we examine the behavior of domains during the early stages of the simulation, i.e. before morphologies are well developed. Initially, close to the starting moment of the overall process, the randomness of the initial distribution of phases is robust. As time elapses, morphologies tend to form, while their precise shape still evolve until stable shapes are reached.  In both the structure factor and correlation function calculations, an oscillating behavior is observed during these early stages in simulations where the morphologies are ill-formed; we emphasize particularly Figure~\ref{fig:Boxes} as well as 
Figure~\ref{fig:SmallTimeStructureFactor}.  This behavior has been observed experimentally as well and is discussed in \cite{Michels_2013_simulation}.  

\subsection{Domain Growth at Large Time Scales}\label{large_scale}
\begin{figure}[htbp]
    \centering
    \includegraphics[scale =0.45]{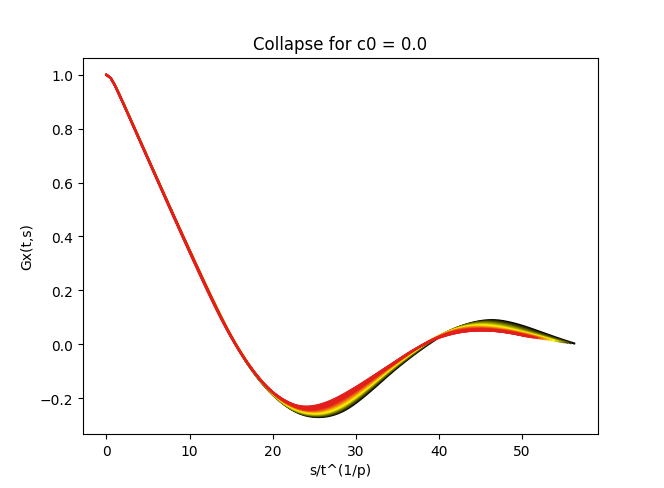}
    \includegraphics[scale =0.45]{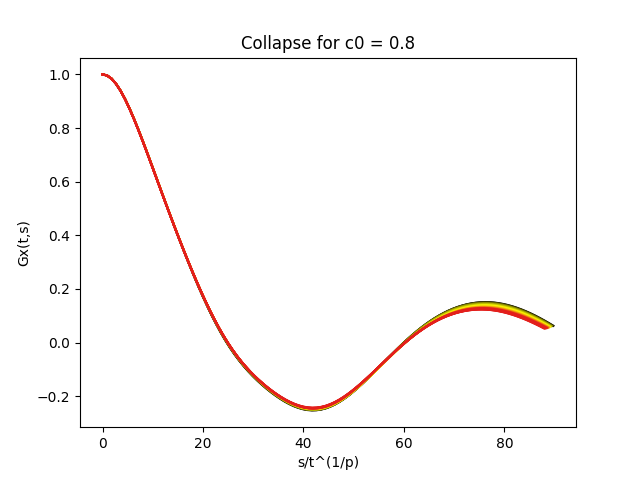}
    \caption{Correlation functions for the time interval $[6,9]$ plotted against the numerical spacial steps scaled according to the power law $t^{1/p}$, where $p= 3$  for $c_0 = 0$ (left) and $p=18$ for $c_0 = 0.8$ (right). The colors of the plots represent the time of the simulation with green representing $t=6$ and red representing $t=9$. All other concentrations are similar with their respective powers.}
    \label{fig_Collapse}
\end{figure}

We examine now the domain growth during a time frame where the morphologies are well formed. In this time frame, we observe from the indicator functions the domains are growing in accordance with power laws. In other words, we observe evidence that the simulations obey the scaling hypothesis \cite{bray2002theory}. 
The accuracy of these power laws is demonstrated 
in Figure~\ref{fig_Collapse}
by rescaling the correlation functions as discussed in the introduction.
The observed behavior for accurate scaling regimes would be a `collapse' of these curves into a single curve in the variable $z:= |s|/t^\alpha $, as shown in Figure~\ref{fig_Collapse}. In other words, this visualization suggests there exists a value of $\alpha$ so that the function $G(t,z)$ is independent of time. We particularly point out from 
Figure~\ref{fig_Collapse} the power 1/3 from the Ising model appears to be close to the true scaling power of the $c_0 =0$ case. This type of demonstration has been used for other phase separation models, see e.g. \cite{bray2002theory,CirilloGonnellaStramaglia1997} and the references within.  

As illustrated in the right-hand plot of Figure~\ref{fig_StructureFactorT9}, the late time behavior of the system at solvent fraction $c_0=0,0.2$ is very much compatible with the typical $1/3$ growth 
exponent characterizing conservative two--state systems dynamics. 
This is confirmed also by the morphologies shown in the first 
two rows of Figure~\ref{fig_m_Morphologies}, where the bicontinuous 
phase typical of Ising--like systems is observed.

The situation is very much different at large values of solvent 
concentration, that is to say for the case $c_0=0.6,0.8$ where the growth 
exponent is much smaller. Correspondingly,  the growth process is much slower, 
and hence, the morphology is characterized by approximately spherical
domains inside a percolating sea of solvent.

Finally, at the intermediate value $c_0=0.4$ of solvent fraction,
a sort of crossover between the small and large solvent fractions 
is observed. Figure~\ref{fig_StructureFactorT9} suggests an exponent 
droping from $1/4$ to $1/12$ around time $t = 3$. The corresponding morphologies are
coherent. Indeed, the initial bicontinuous phase progressively transforms
into a picture characterized by well separated domains inside the 
sea of solvent.

\subsection{Behavior for Small Concentrations}

It is natural to contemplate the behavior of the growth power laws discussed in the previous sections as the limit of the solvent concentration tends to 0. We plot the results of the structure factor and correlation function indicators 
in Figure~\ref{fig:SmallTimeStructureFactor}. We observe that the limiting behavior of the simulation non--monotonically tends toward the behavior of the two state model with $0$ solvent concentration level. Assuming the behavior changes continuously with the initial solvent concentration, this would imply the existence of an optimal initial solvent proportion maximizing the power, $1/p$. We will also point out that since the mesh size $\Delta x , \Delta y$ is fixed for all simulations, there may be issues tracking the solvent at small (relative to the mesh size) concentrations. 

\begin{figure}[htbp]
    \centering
    \includegraphics[scale = 0.5]{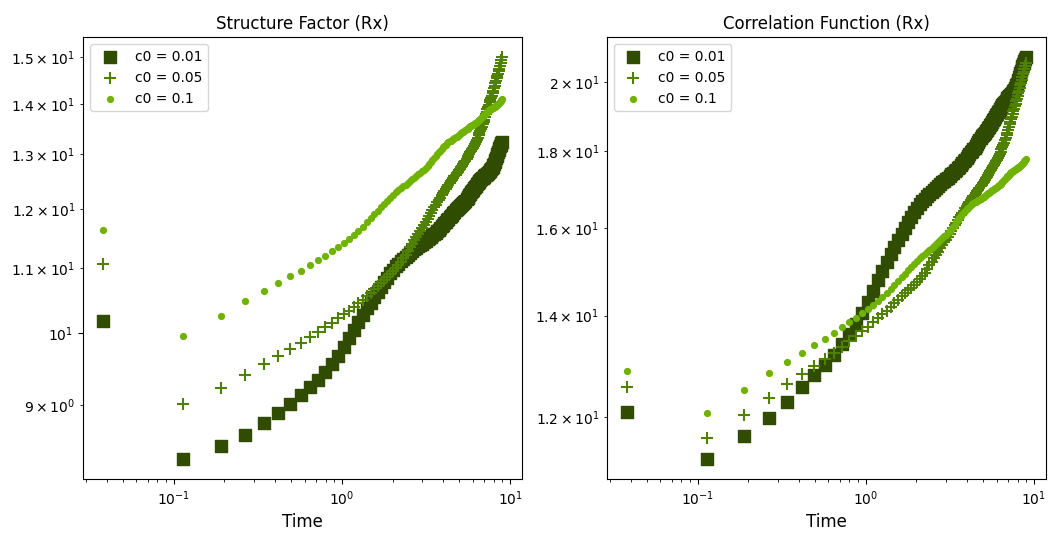}
    \caption{Structure factor (left) and correlation function (right) calculations for solvent concentrations $c_0 = 0.1,0.05,0.01$ during the time interval $(0,9]$. Each marker represents $4,000$ time steps in the finite volume scheme. }
    \label{fig:StructureFactor_SmallConcentration}
\end{figure}

\section{Conclusion and Outlook}\label{outlook}
Taking into consideration the results presented, it is clear that the behavior of model \eqref{Eq_MainModel} is rich with complex dynamics and possesses the aptitude for phase separation. In particular, as the model was derived as the hydrodynamic limit of the Blume--Capel model with Kawasaki dynamics and Kac interactions, it may prove to be a useful continuum model for the Monte--Carlo simulations addressed in \cite{Andrea_PhysRevE,Mario} without evaporation. Of course, the model currently has no machinery to capture the evaporation effect on the solvent particles and there is less flexibility (with respect to the Monte--Carlo simulation) in the tuning of interaction parameters. To mend these differences, potential modifications to the model, including for example the addition of an evaporation boundary or interaction tensor, must be investigated. 

 From a mathematical perspective, the well--posedness and regularity of solutions to this model remains to be studied. While \cite{Marra} provides existence of solutions via a hydrodynamic limiting process and uniqueness in $L^\infty$, the parabolic structure of the equations in \eqref{Eq_MainModel} suggests that more regularity is expected. Once a suitable space for solutions is established, convergence of the finite volume scheme can be rigorously shown. Numerical simulations point out that we can also expect certain properties on the relationship between $m$ and $\phi$ (e.g. $|m| \leq \phi$ for all $t>0$). Showing these relationships hold for model \eqref{Eq_MainModel} may have strong implications for the different applications. 
 
 The analysis on the domain growth done throughout Section \ref{Sec_DomainGrowth} suggests this model satisfies the scaling hypothesis presented in \cite{bray1990universal}. Moreover, it appears that solvent can behave as a type of catalyse in the growth of the morphologies. This behavior has also been observed in the stochastic model \cite{Andrea_PhysRevE}. It remains to be shown if the powers presented in Section \ref{Sec_DomainGrowth} can, under some assumptions, be analytically calculated from model \eqref{Eq_MainModel} as a function of the initial solvent concentration. One may follow some of the arguments in \cite{bray2002theory} to formulate potential ideas for calculating approximate expressions of the  growth laws, i.e. to get insight into the so--called coarsening rates at least for simple ball--like morphologies. It is worth noting that our numerical simulations indicate that for $c_0=0$, the model recovers domain growths as expected in the two--species Ising model in two--space dimensions. As $c_0$ increases up to some value around $0.5$ a percolation regime is noticed, while if $c_0$ takes relatively high values, then the growth of morphologies is slowed down accordingly, as in fact expected from the physical point of view.

Last but not least, the continuum model is superior to its stochastic microscopic description in terms of computation time. This feature keeps open the possibility of fast estimation of  model parameters especially if one thinks of involving more complex physics, like  accounting for solvent evaporation, transport and recombination of charges along the obtained morphologies, polycrystallization, degradation --  typical scenarios encountered, for instance, in the case of organic solar cells; see \cite{Benoit_2021, Harting_ATS} for recent related works.

\section*{Acknowledgments}
 RL and AM are grateful to Carl Tryggers Stiftelse for the financial support via the grant CTS 21--1656. ENMC acknowledges the Mathematics Department of the Karlstad University for its warm hospitality as well as INDAM--GNFM for support. AM thanks SNIC for projects nr. 2020/9--178, 10--94 (HPC2N), and 2022/22-1171 {\em Multiscale simulations of hybrid continuum--discrete--stochastic systems} for providing computational resources and storage capacity. SAM acknowledges the funding from the Swedish National Space Agency (Grant 174/19) and the Knut and Alice Wallenbergs Stiftelse (Grant 2016.0059) 
 
\bibliographystyle{plain}
\bibliography{morpho}
\end{document}